\documentclass[aps,prb,twocolumn,showpacs,showkeys]{revtex4}

\usepackage{psfig,graphics,graphicx}
\usepackage{color}
\usepackage{amssymb}
\usepackage{bookmath}

\definecolor{red}{rgb}{1,0,0}

\begin{document}

\preprint{LA-UR-05-3719}

\title{Phase Diagram and Spectroscopy of FFLO states of two-dimensional $d$-wave superconductors}
\author{A. B. Vorontsov}
\affiliation{Department of Physics and Astronomy, 
Louisiana State University, Baton Rouge, LA 70803,}
\affiliation{Center for Nonlinear Studies,
Los Alamos National Laboratory, Los Alamos, NM 87545, USA.}
\author{J. A. Sauls}
\affiliation{Department of Physics and Astronomy, 
Northwestern University, Evanston, IL 60208,}
\affiliation{Department of Physics, 
University of Florida, Gainesville, FL 32611, USA.}
\author{M. J. Graf}
\affiliation{Theoretical Division, 
Los Alamos National Laboratory, Los Alamos, NM 87545, USA.}

\date{\today}

\begin{abstract}
Experimental evidence suggests that the Fulde-Ferrell-Larkin-Ovchinnikov (FFLO)
state may be realized in the unconventional, heavy-fermion superconductor CeCoIn$_5$.
We present a self-consistent calculation of the field versus temperature phase diagram 
and order parameter structures for the FFLO states of quasi-two-dimensional $d$-wave superconductors. 
We calculate the spatially nonuniform order parameter, free energy density, and local density of states 
for magnetic fields parallel to the superconducting planes. 
We predict that the lower critical magnetic 
field transition between the spatially uniform and nonuniform FFLO state is second order. We discuss 
the signatures of the nonuniform FFLO state which should be observable in scanning tunneling microscopy 
measurements of the local density of states.
\end{abstract}
\maketitle

\section{\label{sec:INTRO} Introduction}

The Fulde-Ferrell-Larkin-Ovchinnikov (FFLO) state is predicted for clean spin-singlet superconductors as 
a result of the competition between pairing correlations, favoring anti-parallel spin alignment, and the 
Zeeman effect, favoring parallel spin alignment along the field.\cite{ful64,lar64} The compromise is 
a spatially inhomogeneous state of ``normal'' and ``superconducting'' regions.
The ``normal'' regions are defined by a spectrum of spin-polarized quasiparticles. 
The high-field FFLO phase was originally suggested for superconductors with ferromagnetically aligned 
impurities,\cite{ful64,lar64} but it was soon realized that a FFLO state should develop in 
superconductors in an external field if the Zeeman coupling dominates the orbital 
coupling.\cite{mak64a,mak66,gru66}

A number of layered organic superconductors have been suggested as candidates for FFLO 
phases.\cite{shi96,sin00,hou02,bul03,tan03,kon04} However, recent interest has focussed
on the heavy-fermion material CeCoIn$_5$.\cite{pet01}
This material has quasi-two-dimensional (2D) metallic planes, and shows evidence of 
unconventional, $d$-wave, superconductivity\cite{mov01,iza01} with a transition 
temperature of $T_c = 2.3\,K$. The quasi-2D tetragonal crystal structure consists of 
alternating of CeIn$_3$ and CoIn$_2$. Superconductivity is believed to develop
on a Fermi surface sheet which is nearly cylindrical, i.e., a sheet in which the 
conduction electrons move primarily in the 2D metallic planes.\cite{hal01}
CeCoIn$_5$ exhibits the necessary characteristics for the existence of the FFLO state.
It is in the clean limit as indicated by the small normal-state residual resistivity 
and the small Sommerfeld coefficient for the low-temperature specific heat in the 
superconducting state.\cite{mov01}
The upper critical field is paramagnetically limited, as indicated by the large ratio 
of orbital critical field to Pauli critical field,\cite{bia02} 
and CeCoIn$_5$ exhibits a critical point where the transition from the normal to superconducting 
state changes from second-order to first order below the temperature $T_0 = 0.7$ or $1.3\,K$, 
depending on the field orientation.\cite{iza01,tay02,tak02,bia02,mur02,bia03} 
Finally, several different experiments indicate a second-order phase transition within 
the superconducting state, which is assumed to be the transition into the FFLO 
state.\cite{rad03,bia03,cap04,wat04,mar05,kak05}

For $s$-wave superconductors it has been predicted that the upper critical field 
of the FFLO state depends on the dimensionality of the Fermi surface,\cite{buz97}
Fermi liquid interactions,\cite{bur94} orbital effects and the presence of 
vortices,\cite{tac96,bul03} and impurities.\cite{mak66,asl68,bul76}
The one-dimensional stripe pattern for the order parameter originally proposed 
by Fulde and Ferrell, and Larkin and Ovchinnikov has been generalized to include
higher harmonics and more complicated 2D FFLO lattices.\cite{buz97,mor04,shi98} 
Burkhardt and Rainer included Fermi liquid effects and examined self-consistently 
quasi-2D superconductors,\cite{bur94} while Matsuo et al.\ formulated their analysis for 
three-dimensional (3D) Fermi surfaces.\cite{mat98} In 2D the normal to superconducting 
transition at $B_{c2}$ is second order, unless Fermi liquid effects are large.
In 3D the FFLO transition is first order at high temperatures, but second order at 
sufficiently low temperatures.\cite{bur94} For both 2D and 3D cases, the transition 
at the lower critical field, $B_{c1}$, from the uniform superconducting state to 
the non-uniform FFLO state is second order and signaled by the appearance of a 
single domain wall. Generally both spin and orbital coupling to the field are 
present. This leads to coexistence of the FFLO with vortices; these states have been 
studied extensively and it is predicted that the resulting inhomogeneous state
has a number of novel properties, including a change in the order of the transition 
as well as additional phase transitions within the nonuniform phase corresponding to
transitions between states belonging to different Landau 
levels.\cite{buz96a,buz96b,tac96,shi97,kle00,hou01,bul03,yan04,kle04}

For $d$-wave superconductors the FFLO phases are modified by the intrinsic anisotropy of 
the order parameter. In the class of quasi-2D superconductors the upper critical 
transition line, $B_{c2}(T)$, has a kink at low temperatures corresponding to the 
discontinuous change in direction of the stripe modulation with respect to the 
nodes of the order parameter.\cite{mak96,shi97,shi98,yan98} 
However, the re-orientation transition, and the kink, are absent in 
3D,\cite{sam96} or if the shape of the Fermi surface is changed.\cite{shi02} 
Recent calculations of the spatial modulation of the order parameter in 2D near $B_{c2}$
predict that the energetically favored state at low temperature and high field 
forms a ``square lattice'' instead of stripe order.\cite{shi98,mak02} The interplay
between Pauli paramagnetism and the vortex phase of $d$-wave superconductors
has also been studied in Refs.~(\onlinecite{shi97,ada03}). 

Theoretical predictions for the properties of the FFLO phases depend on material 
properties, including the quasiparticle interactions. Accurate results for the 
thermodynamic and transport properties of the FFLO state generally require self-consistent 
calculations of the inhomogeneous order parameter structure, local excitation spectrum 
and quasiparticle self energy. This is often a formidable task as one must examine 
a great number of different, but energetically nearby, states.
Most of the theoretical work mentioned above was limited to the vicinity of the upper 
critical field, or near the maximum critical transition temperature, 
$T_\sm{FFLO}$, of the FFLO state.  

In this paper we address the stability of the nonuniform FFLO phases 
of 2D $d$-wave superconductors over the field range from the lower critical field, $B_{c1}$,
to the the upper critical field, $B_{c2}$. Our analysis is for applied fields parallel to the 
superconducting layers. In this geometry the effect of the magnetic field on the superconducting 
condensate enters only through the Zeeman coupling of the quasiparticle spin to the field.
For simplicity we assume a cylindrical Fermi surface, which is also supported by de Haas-van Alphen 
measurements on CeCoIn$_5$.\cite{hal01}

The outline of this paper is as follows:
In Sec.~\ref{sec:THEO} we introduce the theoretical model and the relevant aspects of 
the quasiclassical formulation of the equations for inhomogeneous superconductors.
The B-T phase diagram for two-dimensional $d$-wave superconductors is discussed in 
Sec.~\ref{sec:PD}. This is followed by Sec.~\ref{sec:DOS} where we report calculations of
the local density of states in the FFLO phases. We examine the energetics of the stripe
phases and address the question of how to observe the spatial modulation of the order 
parameter structures in Sec.~\ref{sec:FE}. Conclusions are provided in Sec.~\ref{sec:CLS}.

\section{\label{sec:THEO} Theoretical model}

The quasiclassical theory of superconductivity\cite{eil68,lar68,eli71} 
is well suited for the study of spatially inhomogeneous states of superconductors
varying over distances large compared to the Fermi wavelength, e.g., the superconducting 
coherence length, $\xi_0=v_f/(2\pi T_c) \gg k_f^{-1}$. We consider spin-singlet superconductors
for which the order parameter factorizes, $\Delta(\vR,\hat{\vp})=\Delta(\vR)\,\cY(\hat{\vp})$,
into a spatially dependent complex amplitude, $\Delta(\vR)$, and a momentum-space basis 
function, $\cY(\hat{\vp})$, for an even-parity, one-dimensional representation of the crystal
point group, where $\hat{\vp}$ is the direction of the Fermi momentum.
The basis function is normalized, $ \langle \cY^2(\hat{\vp}) \rangle_\sm{FS} = 1$, where
$\langle\ldots\rangle_\sm{FS}$ represents the angular average over the Fermi surface
weighted by the relative angle-resolved normal-state density of states on the Fermi surface.

In the following we consider the FFLO states of a clean $d$-wave superconductor with
$d_{x^2-y^2}$ symmetry described by the one-dimensional basis function,
$\cY(\hat{\vp}) = \sqrt{2}\,(\hat{p}_x^2 - \hat{p}_y^2) = \sqrt{2}\,\cos 2\phi_{\hat{\vp}}$.
In this case the magnitude of the zero temperature gap parameter is $\Delta_{0}/(2\pi T_c) = 
0.241$.\cite{footnote}

We solve the Eilenberger equation in a magnetic field $\vB$ for the quasiclassical (Matsubara) 
Green's function $\whg(\vR, \hat{\vp}; \vare_m)$, 
\be
\label{eq:eil}
[ i\vare_m \widehat{\tau}_3 - \whDelta - \widehat{v} , \whg ] 
+ i\vv_f \cdot \grad \, \whg = 0 \,,
\ee
self-consistently with the order parameter, $\Delta(\vR)$.
In addition, the Green's function must satisfy the normalization condition
\be
\whg^2 = -\pi^2 \widehat{1} .
\ee
The quasiclassical Green's function in Nambu space 
can be decomposed into scalar and spin vector components
\be
\whg = \left(
\begin{array}{cc}
g + \vg \cdot \vsigma & (f+\vf \cdot \vsigma)i\sigma_2 \\
i\sigma_2 (f'+\vf' \cdot \vsigma) & -g + \vg \cdot \vsigma^*
\end{array}
\right) 
\,.
\ee
Correspondingly, the order parameter matrix in Nambu space for a spin-singlet superconductor is
\be
\whDelta(\vR, \hat{\vp}) = \left(
\begin{array}{cc}
0 & i\sigma_2\, \Delta \\
i\sigma_2\, \Delta^* & 0 
\end{array}
\right) \,.
\ee
The Zeeman coupling of the quasiparticle spin with magnetic field is given by
\be
\widehat{v}= + \left(
\begin{array}{cc}
\mu \vB \cdot \vsigma & 0 \\
0 & \mu \vB \cdot \vsigma^*
\end{array}
\right) \,,
\ee
where $\sigma_i$ are Pauli spin matrices and $\mu = (g/2)\mu_B$ is 
the absolute value of the magnetic moment of a quasiparticle with negative charge $e$; 
$\mu_B=|e|/(2mc)$ is the Bohr magneton. Note that the $g$-factor 
is a material parameter, as is the Fermi velocity, which incorporates 
the high-energy, short-wavelength renormalizations of the bare electron g-factor into the 
effective g-factor defining the Zeeman coupling of the quasiparticle to the field. 

\begin{figure}[t]
\centerline{\includegraphics[width=8.0cm]{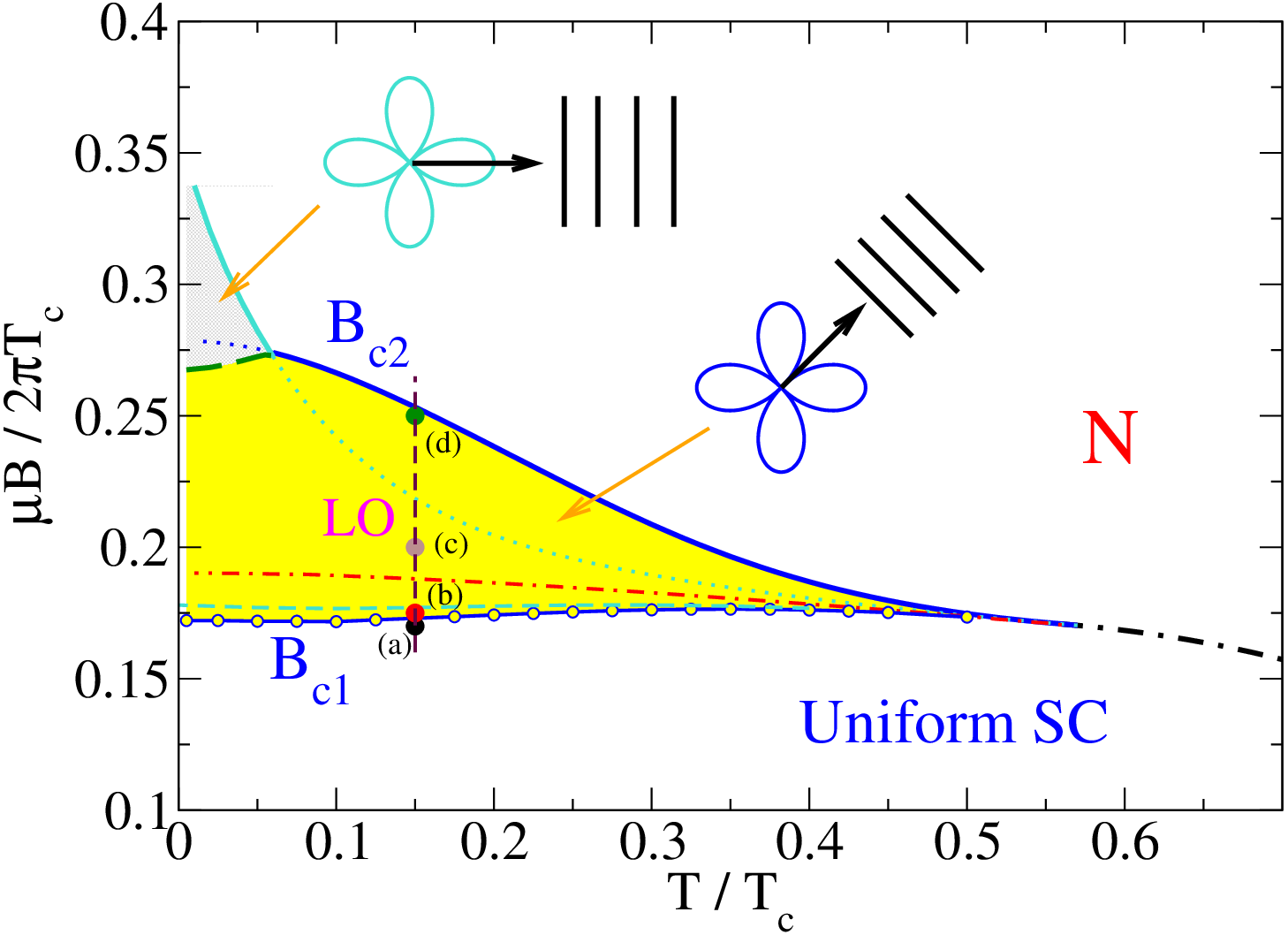}}
\caption{\label{fig:PD} 
The phase diagram of the FFLO state of a two-dimensional 
$d$-wave superconductor. The Larkin-Ovchinnikov (LO) state is 
stabilized in the high-$B$, low-$T$ region of the phase diagram. 
The solid lines are second order phase transition lines that determine the 
upper critical field, $B_{c2}$, and separate the normal and FFLO states
below $T_\sm{FFLO}\approx 0.56\, T_c$.
Below $T \lesssim 0.06\,T_c$ a first order transition (long-dashed line) 
occurs between order parameter modulations along $\langle 110 \rangle$ and
$\langle 100 \rangle$ directions.
At the lower critical field, $B_{c1}$, a second order transition (circles-solid) 
occurs between the uniform and nonuniform $\langle 110 \rangle$-oriented LO phase.
The unphysical transition line from the uniform state into the 
$\langle 100 \rangle$-oriented nonuniform state is shown for comparison (short-dashed). 
The Chandrasekhar-Clogston phase transition line between the uniform superconducting 
and normal state would be of first order below $T_\sm{FFLO}$  (dot-dashed),
but is unphysical, while it is of second order and physical above $T_\sm{FFLO}$ 
(thick dot-dashed).
The corresponding order parameter modulations for a field scan between points 
(a) through (d) are shown in Fig.~(\ref{fig:op}).
}
\end{figure}
\begin{figure}[t]
\centerline{\includegraphics[width=8.0cm]{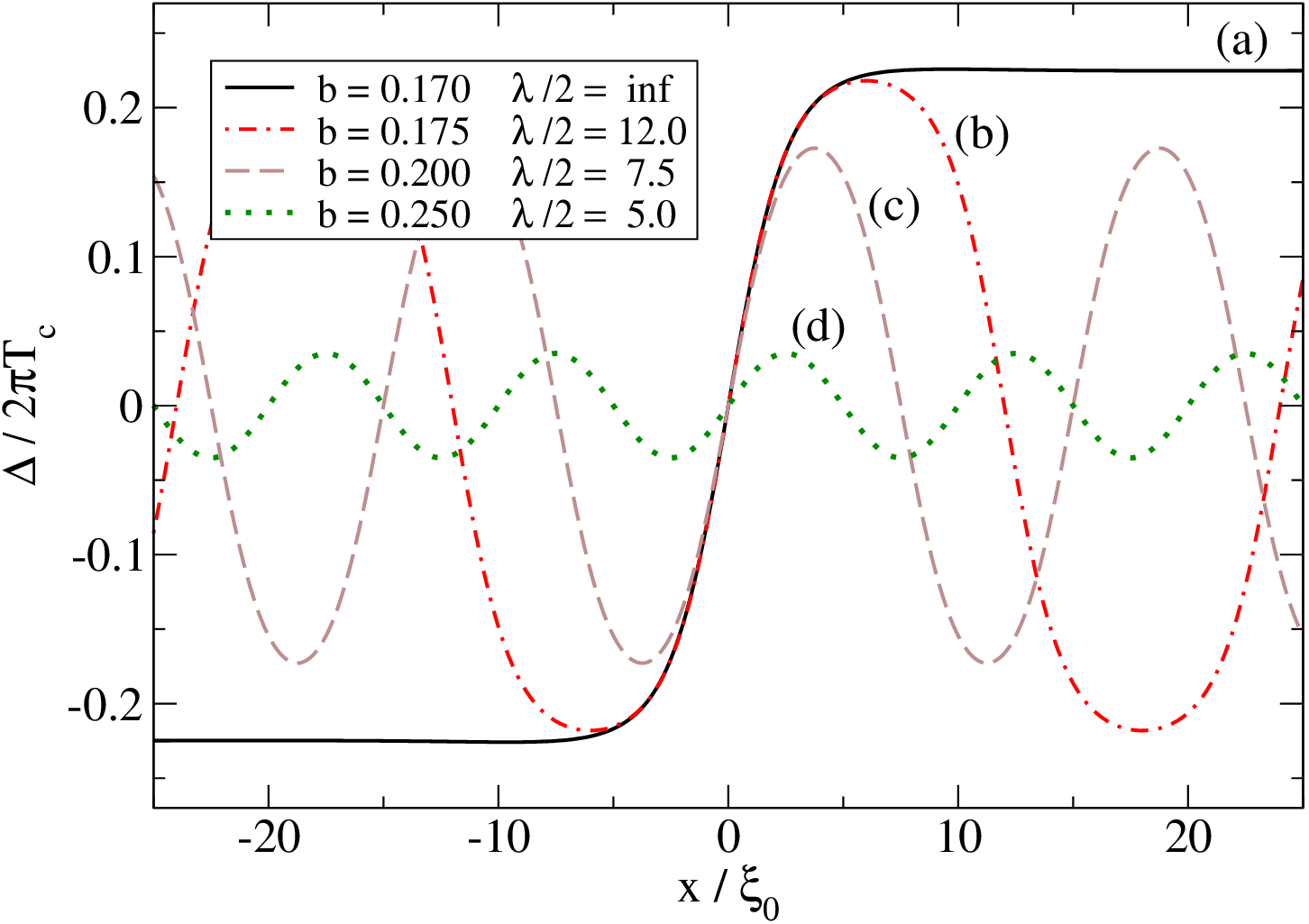}}
\caption{\label{fig:op} 
Order parameter modulations at $T = 0.15 \, T_c$ 
for different points in the LO state as indicated by (a)-(d) in Fig.~(\ref{fig:PD}). 
The period of the order parameter decreases with increasing field $B$. 
At the lower magnetic field, $B_{c1}$, a transition from the
uniform to nonuniform superconducting state occurs at position (a), 
which is signaled by a single domain wall (kink). 
We define the superconducting coherence length $\xi_0 = v_f /(2\pi T_c)$ 
and dimensionless magnetic field $b=\mu B/(2\pi T_c)$, 
with the zero-temperature order parameter 
$\Delta_{0}/(2\pi T_c)  = 0.241$. 
}
\end{figure}

For a spatially uniform order parameter, $\Delta = \Delta(\hat{\vp})$,
the physical solution of the Eilenberger equation
for the Green's function $\whg(\hat{\vp}; \vare_m)$ is
\bea 
&&
g = -{\pi\over 2} \left( \frac{i\vare_+}{D_+} + \frac{i\vare_-}{D_-} \right) ,\,
\vg = -{\pi\over 2} \left( \frac{i\vare_+}{D_+} - \frac{i\vare_-}{D_-} \right) \hat{\vb},
\nonumber \\
&&f = {\pi\over 2} \Delta \left( \frac{1}{D_+} + \frac{1}{D_-} \right) ,\,
f' = {\pi\over 2} \Delta^* \left( \frac{1}{D_+} + \frac{1}{D_-} \right), 
\nonumber \\
&&\vf = {\pi\over 2} \Delta \left( \frac{1}{D_+} - \frac{1}{D_-} \right) \hat{\vb} ,\,
\vf' = {\pi\over 2} \Delta^* \left( \frac{1}{D_+} - \frac{1}{D_-} \right) \hat{\vb} \,,
\nonumber 
\eea
where
\bea 
\label{eq:energy}
i\vare_\pm = i\vare_m \mp \epsilon_\sm{B} \quad,\quad \epsilon_\sm{B} = \mu B \,, \\
D_\pm(\hat{\vp}) = \sqrt{|\Delta(\hat{\vp})|^2 -(i \vare_\pm)^2} \,,
\eea
and $\hat{\vb}$ is the direction of the uniform magnetic field.

The uniform solution is easily generalized to the Fulde-Ferrell (FF) state
in which the order parameter has the spatially varying phase,
$\Delta(\vR, \hat{\vp}) = \Delta(\hat{\vp})\, \exp( i\vq\cdot\vR )$.
The solution in this case is obtained by replacing the frequency in 
Eq.~(\ref{eq:energy}) by $ i\vare_\pm \to i\vare_\pm - \eta$, 
where $\eta = {1\over 2}\vv_f\cdot\vq$. The Green's function is then obtained 
from the uniform solution by a uniform gauge transformation 
$\whg(\vR,\hat{\vp}; \vare_m) = \widehat{U} \, 
 \whg(\hat{\vp}; \vare_m)\, \widehat{U}^\dag$ 
where $\widehat{U} = \exp (i\, \vq\cdot\vR \, \widehat{\tau}_3/2)$.
The linearized form of these solutions are used to obtain the upper critical field, 
$B_{c2}(T)$, for the FFLO states, as well as the order parameter and thermodynamic
functions for the homogeneous phase below $B_{c1}(T)$.

Obtaining the Green's function for a nonuniform order parameter 
is more challenging, and generally requires numerical solution of Eilenberger's
equation. Schopohl\cite{sch95a,sch96} and Nagai\cite{nag93,nag96b} 
introduced an efficient and numerically stable method for solving Eilenberger's 
equation by transforming it to a Riccati-type equation. We use the Riccati
method as described by Eschrig.\cite{esc00} From the solutions to the Riccati
equation we can construct the components of the Green's function. The diagonal 
Green's functions, $g$ and $\vg$, determine the quasiparticle excitation spectrum,
while the off-diagonal Green's functions, $f$ and $\vf$, determine the order parameter 
through the BCS gap equation. For pure spin-singlet $d$-wave pairing, 
\bea
\lefteqn{
\Delta(\vR) \ln {T\over T_c} = 
T\, \sum_{\vare_m > 0} \int \done{\hat{\vp}} \; \cY(\hat{\vp}) \; \times 
}&&
\label{eq:sc}
\\ 
&& \left(f(\vR, \hat{\vp}; \vare_m) + f'(\vR, \hat{\vp}; \vare_m)^* - 
2{\pi \Delta(\vR) \cY(\hat{\vp}) \over |\vare_m|} \right) \,.
\nonumber
\eea
The gap equation is solved self-consistently together with the Riccati equations.

There are often many nearly degenerate order parameter configurations. 
In order to examine the relative stability of different order parameter solutions,
we also calculate the free energy obtained from a quasiclassical reduction of the 
Luttinger-Ward functional,\cite{ser83,vor03c} 
\be
F_S(B,T) - F_N(B,T) = \int d\vR \, \Del f(\vR) \,,
\ee
where the difference functional is
\be
\Del f(\vR) = {1\over 2} \int_0^1 \, d\lambda \; 
T\sum_{\vare_m} N_\sm{F} \int \done{\hat{\vp}} \, \mbox{Tr} \; 
\whDelta \left(\whg_\lambda - {1\over 2} \whg \right)  
\,.
\ee
$N_\sm{F}$ is the density of states per spin at the Fermi level of the normal state.
Note that $\whg_\lambda$ is an auxiliary propagator obtained from the solution to the
transport equation with the physical order parameter scaled by the dimensionless coupling 
parameter, $0<\lambda<1$,
\be
[ i\vare_m \widehat{\tau}_3 - \lambda \whDelta - \widehat{v} , \whg_\lambda ] 
+ i\vv_f \cdot \grad \, \whg_\lambda = 0 
\,.
\ee

\begin{figure}[t]
\centerline{\includegraphics[width=8.0cm]{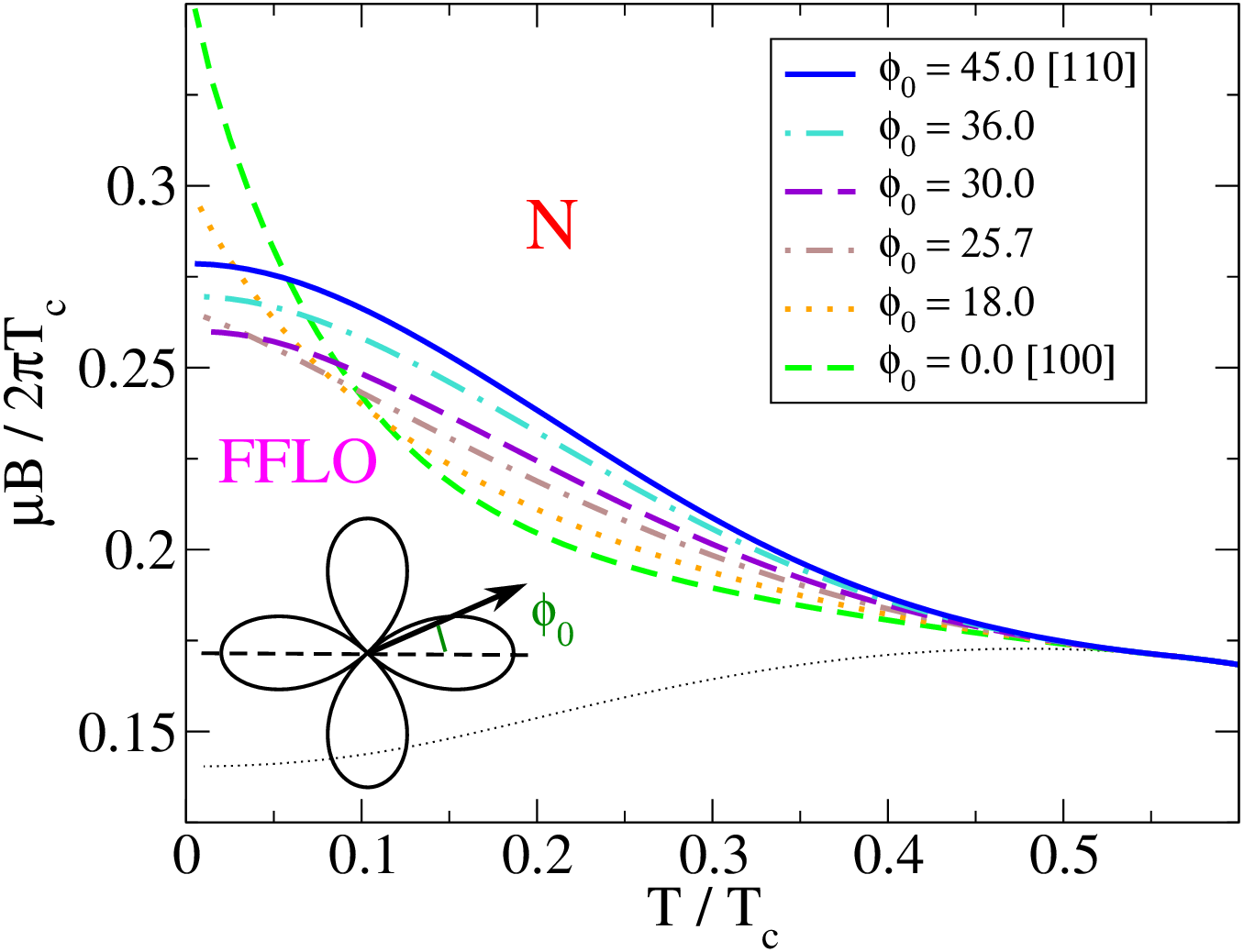}}
\caption{\label{fig:FFLO-N} 
Second order instability lines between normal and FFLO state as function of 
relative orientation $\phi_0$ of one-dimensional stripes, 
i.e., order parameter modulations. 
For temperatures above $T^*\approx 0.06 \,T_c$, the modulation
along $\langle 110 \rangle$
has the highest instability field, while below this temperature the
orientation $\langle 100 \rangle$ is energetically favored. 
The thin-dotted line is the unphysical critical field, or
``supercooled transition'', if one assumes 
a second order transition between the normal and
uniform ($\vq=0$) superconducting state.
}
\end{figure}

\begin{figure}[t]
\centerline{\includegraphics[width=8.0cm]{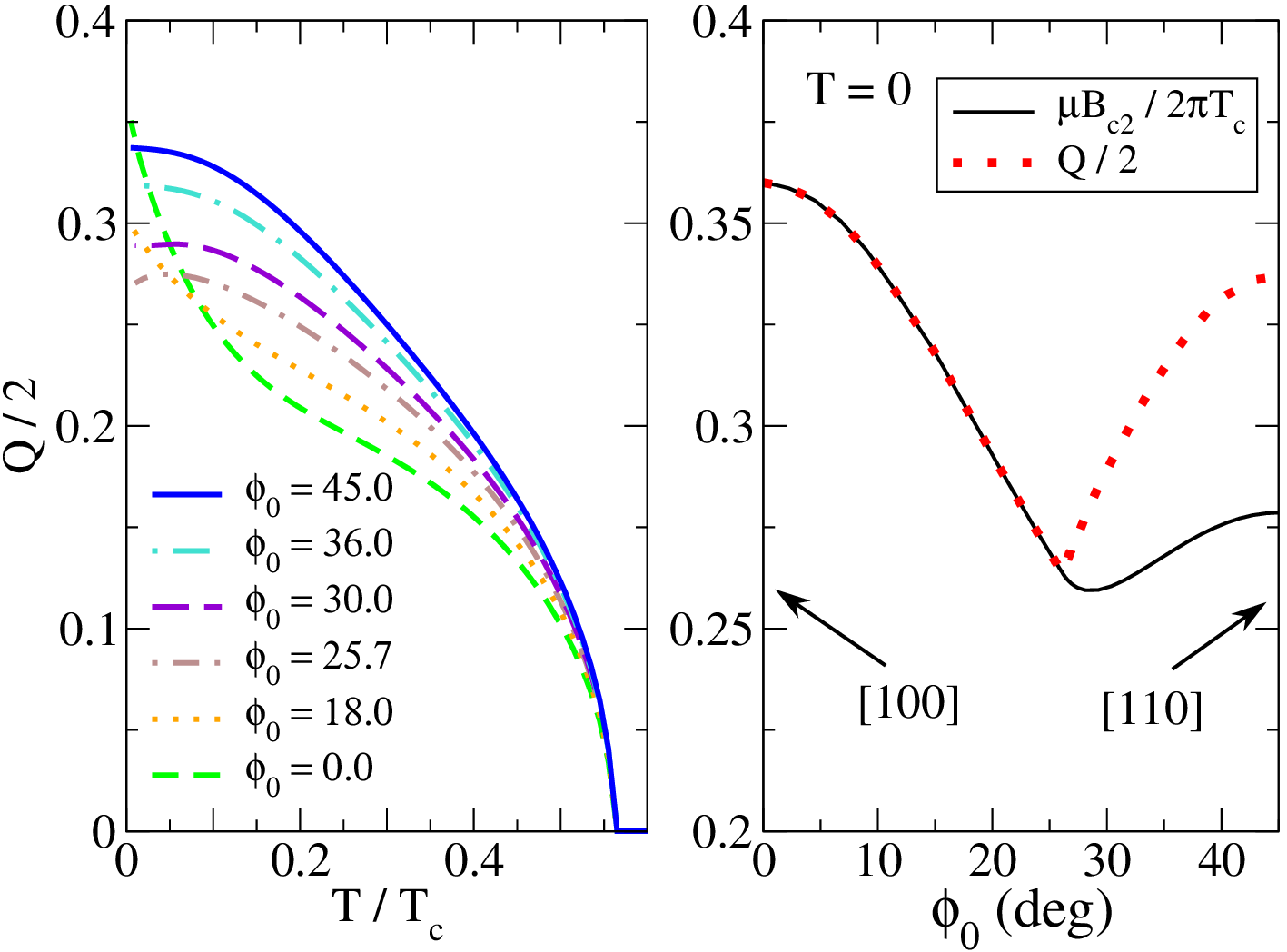}}
\caption{\label{fig:zeroT_BQ} 
Dimensionless wave vector $Q = q\, \xi_0$ of order parameter modulation along the 
instability lines in Fig.~(\ref{fig:FFLO-N}).
Left panel: $T$-dependence of $Q$ for different orientations, with 
$\phi_0 = 0^o \leftrightarrow \langle 100 \rangle$ and 
$\phi_0 = 45^o \leftrightarrow \langle 110 \rangle$.
Right panel: Angular dependence of $Q$ (dots) and upper critical field $B_{c2}$
(solid line) at $T=0$.
}
\end{figure}

The zero-temperature value of the free energy of a uniform 2D $d$-wave superconductor 
differs from the one of an $s$-wave superconductor not only by the angular Fermi surface
averages, but also by an extra term due to the nodal regions of the order parameter,
\bea
\label{eq:F_0}
\Del F(T=0)&&= \; 
-N_\sm{F} \left( \frac{\langle |\Delta(\hat{\vp})|^2\rangle_\sm{FS}}{2}-\epsilon_\sm{B}^2 \right) 
\\ &&
- N_\sm{F} \langle \epsilon_\sm{B} \sqrt{\epsilon_\sm{B}^2-|\Delta(\hat{\vp})|^2} \; 
\Theta (\epsilon_\sm{B}^2-|\Delta(\hat{\vp})|^2) \rangle_\sm{FS} \,.
\nonumber 
\eea
Consequently, the Pauli limiting field of an isotropic $s$-wave superconductor
is $\mu B_\sm{P} = 1.25\, T_c$.  While for a $d$-wave superconductor 
$\mu B_\sm{P} = 1.19\, T_c$, due to the nodal regions on the Fermi surface and in 
particular due to the second bracket in Eq.~(\ref{eq:F_0}).

\section{\label{sec:PD} Phase diagram}

Various aspects of the phase diagram for the FFLO states of a 2D $d$-wave superconductor,
particularly the region near $B_{c2}$, are discussed in the literature.
We extend those studies down to the lower critical field, $B_{c1}$. We consider one-dimensional 
order parameter modulations resulting in one-dimensional stripes. 
Neglecting 2D modulations could result in a modified phase diagram at ultra-high fields 
and ultra-low temperatures.\cite{shi98} 

The form of the order parameter amplitude for the FFLO states is then
\be
\Delta(x,\hat{\vp}) = \Delta(\hat{\vp}) \, f(x) \,,
\label{eq:modelOP}
\ee
where $f(x+\lambda) = f(x)$ is periodic with $\lambda = 2\pi/q$, and 
$x$ is the projected distance along $\vq$; $x = \vq \cdot \vR/q$.
The function $f(x)$ can be either real or complex and 
is a generalization of the Fulde-Ferrell state with a complex 
order parameter solution, 
\be
\Delta_\sm{FF}(\vR, \hat{\vp}) = \Delta(\hat{\vp}) \, e^{i\,\vq\cdot\vR} \,,
\ee
or the Larkin-Ovchinnikov state with a real solution, 
\be
\Delta_\sm{LO}(\vR, \hat{\vp}) = \Delta(\hat{\vp}) \, \sin (\vq\cdot\vR) \,.
\ee
In the vicinity of $B_{c2}$, the gap equation Eq.~(\ref{eq:sc}) can be linearized, 
if one assumes that the transition into the normal state is of second order.
Therefore the LO state, whose order parameter solution is a linear superposition 
of FF solutions with wave vectors $\pm \vq$, has the same instability line 
$B_{c2}$ as the FF state.
 
The wave vector $\vq$, or the period $\lambda$, are implicit functions of temperature 
and magnetic field. In the $s$-wave case all directions are equivalent and 
the modulation, $\vq$, can point along any direction in real space.
Thus, the translational invariance is broken spontaneously in the FFLO state.
In a $d$-wave superconductor the intrinsic anisotropy of the order 
parameter in momentum space favors specific directions for the broken translational symmetry.
When $\vq$ points along the nodal direction of $\cY(\hat{\vp})$, 
we refer to this FFLO phase as the $\langle 110 \rangle$ stripe orientation, and when
$\vq$ is along the anti-nodal direction it is the $\langle 100 \rangle$ orientation.

Finally, in order to calculate the phase diagram we need to compute 
self-consistently the order parameter and find the free energy as a function of
the triplet of parameters, $T,B, \lambda$. The physical FFLO state 
that is realized at a given temperature and field, $(T,B)$,
corresponds to a period $\lambda$ that minimizes the free energy. 
The resulting phase diagram of FFLO states is shown in Fig.~(\ref{fig:PD}).

\begin{figure}[t]
\centerline{\includegraphics[width=6.0cm]{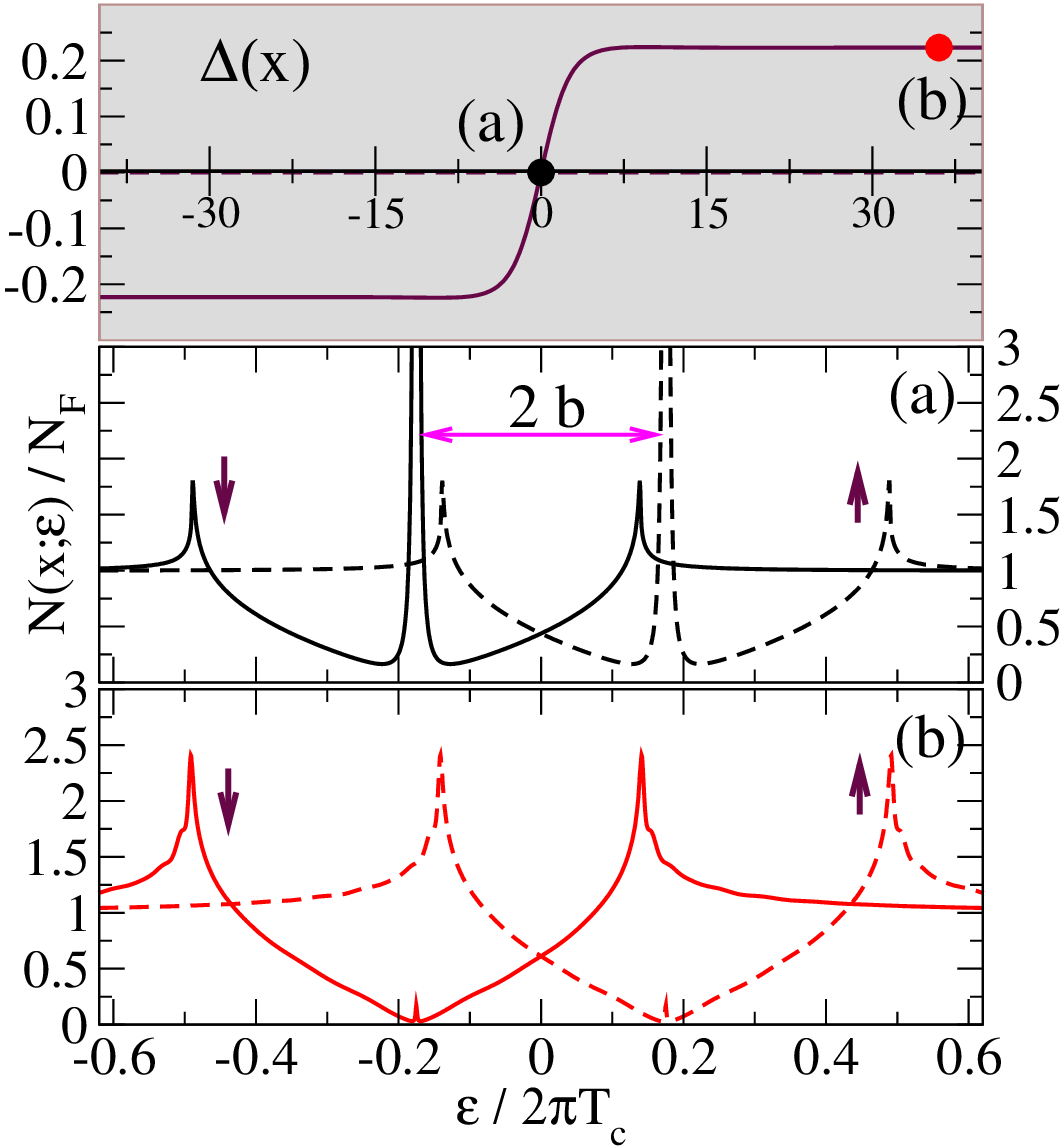}}
\caption{\label{fig:dos110sd} 
Angle-averaged, spin-dependent local density of states (LDOS)
for a single domain wall solution. 
(Top): Profile of order parameter at $T/T_c = 0.15$ and $b=0.175$. 
The panels below show LDOS at locations (a) and (b). 
(Center): LDOS at position (a), where order parameter is zero, i.e., 
at domain wall. The Andreev bound states are shifted from zero energy 
by $+b$ for spin-up and $-b$ for spin-down electrons. 
(Bottom): LDOS at position (b), far from the domain wall, i.e.,
in bulk of $d$-wave superconductor. 
The Andreev states are bound to the region where the order parameter 
is suppressed and are very weak far from the domain wall.  
}
\end{figure}
\begin{figure}[t]
\centerline{\includegraphics[width=6.0cm]{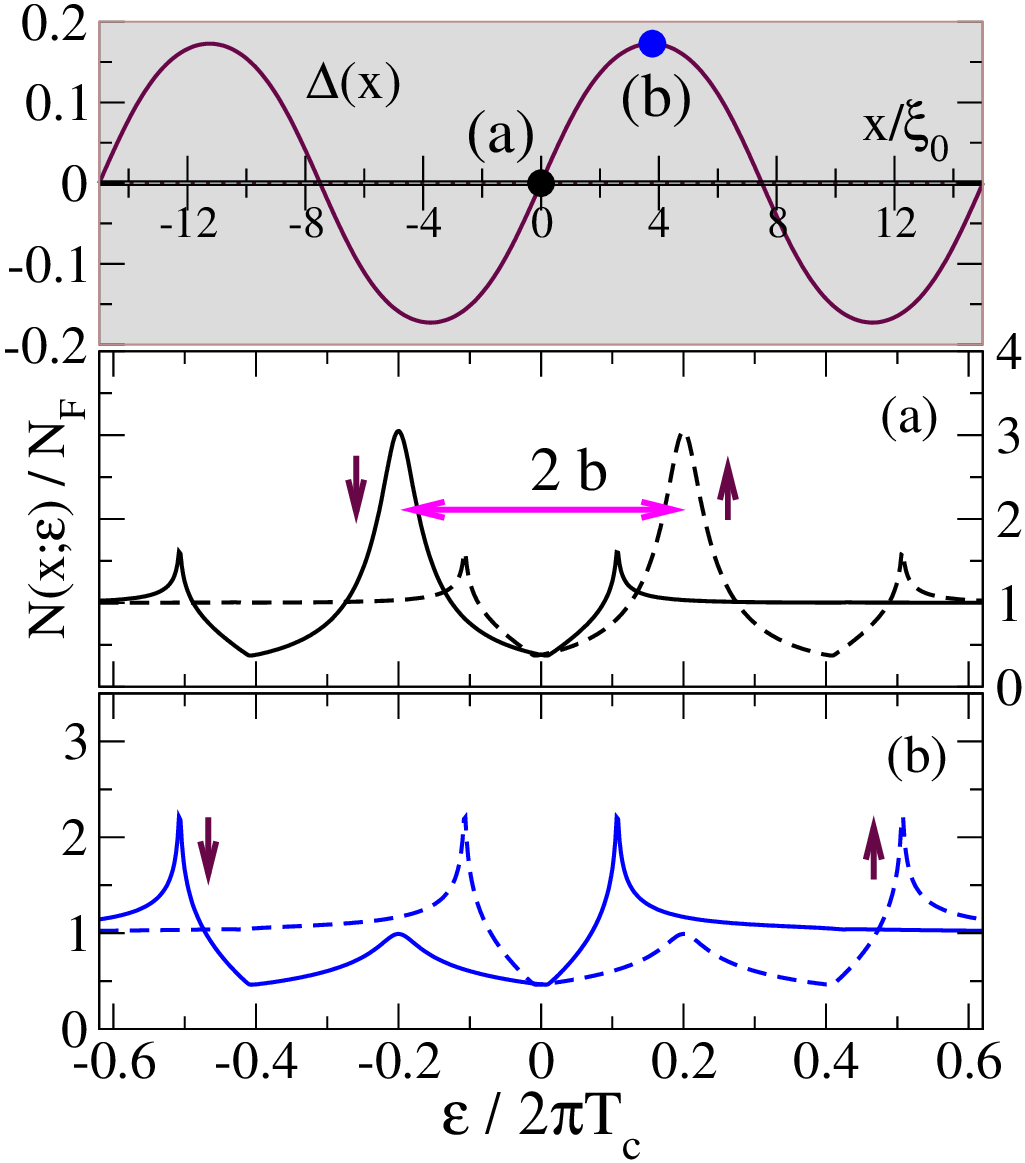}}
\caption{\label{fig:dos110ps} 
Angle-averaged LDOS for a periodic solution inside the LO state. 
(Top): Profile of order parameter at $T/T_c = 0.15$ and $b=0.200$. 
The panels below show LDOS at locations (a) and (b). 
(Center): LDOS at position (a), where order parameter vanishes.
The broadened Andreev bound states, quasiparticle resonances,
are shifted from zero energy by $+b$ for spin-up and $-b$ for spin-down electrons. 
(Bottom): LDOS at position (b) at maximum of order parameter.
The Andreev bound states decayed and broadened even more.
}
\end{figure}

We find that the real (Larkin-Ovchinnikov) order parameter
is favored over the complex (Fulde-Ferrell) order parameter everywhere 
in the nonuniform FFLO region of the phase diagram. The upper critical fields, $B_{c2}$,
for different stripe orientations shown in Fig.~\ref{fig:PD}
(solid and continuing into dotted lines) cross at $T^* \approx 0.06\, T_c$, resulting 
in a kink in the upper transition
line, in agreement with previous work.\cite{mak96,shi97,shi98,yan98} 
On the other hand, the lower critical field, $B_{c1}$, for the stripe orientation 
along $\langle 110 \rangle$ (line with circles) is always 
lower than the critical field for the stripe orientation along 
$\langle 100 \rangle$ (short-dashed line). 
This indicates a first order transition between states 
with different orientation of stripes (long-dashed line). 
Except for a small region in the FFLO phase diagram at ultra-high 
fields and ultra-low temperatures, where the $\langle 100 \rangle$ phase
becomes favorable, the spatial modulation along the $\langle 110 \rangle$ direction
is more stable.

The order parameter profiles for a series of magnetic fields in the LO phase are shown in 
Fig.~\ref{fig:op}. As in the $s$-wave case, the period of the FFLO state decreases with 
increasing magnetic field. The transition between the uniform superconducting and FFLO states
is second order and is signaled by the appearance of a single domain wall. 
The lower critical field transition is analogous to the $s$-wave scenario,\cite{bur94,mat98} 
which contradicts the claim of a first order transition by Yang and Sondhy.\cite{yan98}
The upper critical field transition from the FFLO to normal state is also 
second order with the amplitude of the order parameter vanishing continuously 
with increasing magnetic field.

Figure \ref{fig:FFLO-N} shows the second order instability line from the 
normal state into the nonuniform FFLO state for different orientations of 
the stripes with respect to the anti-nodal direction, denoted by the angle $\phi_0$.
This transition is obtained by linearizing Eqs.~(\ref{eq:eil}) and (\ref{eq:sc}) 
with respect to the order parameter 
$\Delta(\vR,\hat{\vp}) = \Delta_0 \, \cY(\hat{\vp})\, \exp(i \,\vq\cdot\vR)$. 
One of two special directions of $\vq$ for the order parameter modulation,
i.e., $\langle 100 \rangle$ and $\langle 110 \rangle$, yields
the highest instability fields for all temperatures. 

In the left panel of Fig.~\ref{fig:zeroT_BQ} the dimensionless 
instability wave vector $Q = q\,\xi_0$ of the nonuniform LO state 
is plotted as a function of temperature for the different transitions shown 
in Fig.~\ref{fig:FFLO-N}. The right panel of Fig~\ref{fig:zeroT_BQ} shows the zero-temperature 
variation of the critical field $B_{c2}$ and wave vector $q$ as a function of the
stripe orientation $\phi_0$. The evolution of the upper critical field and instability wave vector
with stripe orientation and temperature is the basis for the phase diagram
shown in Fig.~\ref{fig:PD}.

\section{\label{sec:DOS} Quasiparticle density of states}

\begin{figure}[t]
\centerline{\includegraphics[width=5.0cm]{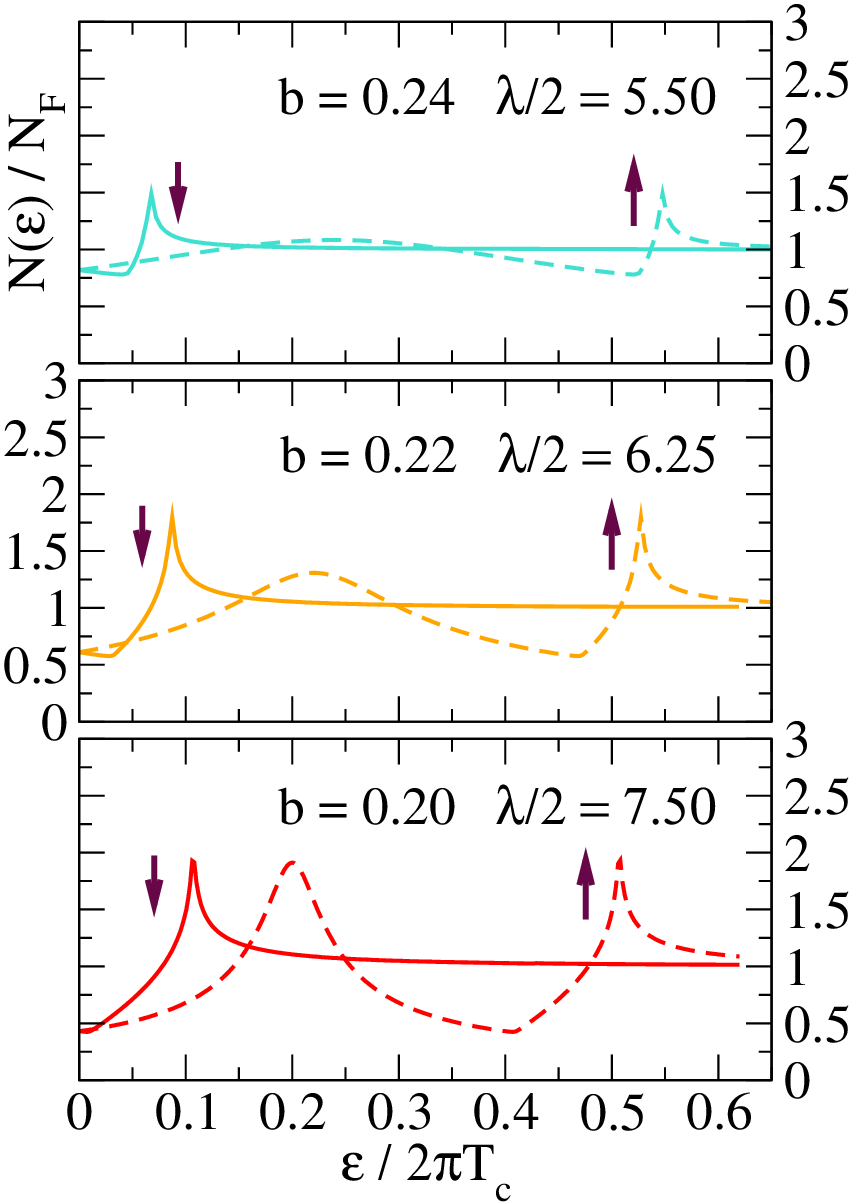}}
\caption{\label{fig:dos110av} 
Spin-up/down DOS averaged over a single period of the order parameter
for three different $b$ fields in the LO state. 
A broad bound state at $\vare/(2\pi T_c)=b$ is seen at lower fields, but 
continues to broaden as $b$ increases. Finally, the DOS becomes flat 
and ``normal''-like.  Here $T/T_c=0.15$ and the corresponding 
upper critical field is $b_{c2}=0.25$.
}
\end{figure}

\begin{figure}[t]
\centerline{\includegraphics[width=8.0cm]{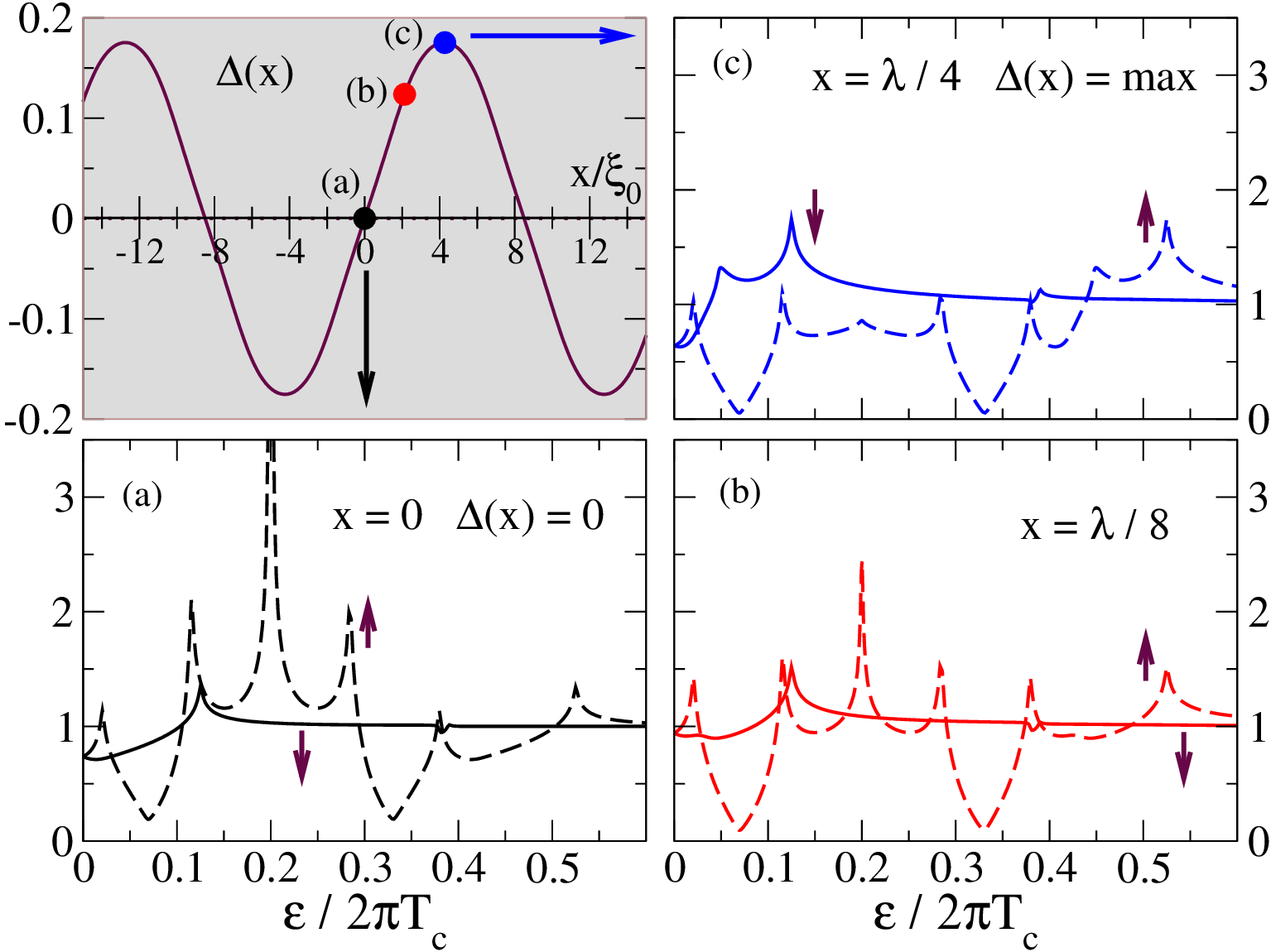}}
\caption{\label{fig:dos100ps} 
An example of a spatially averaged DOS for $\langle 100 \rangle$ modulated 
order parameter at $T = 0.15\,T_c$ and $b = 0.200$. Note that the physically stable
solution has $\langle 110 \rangle$ modulation.
(Upper left): Profile of periodic order parameter with three marked positions.
(Lower left): DOS at the domain wall ($x=0$), the Andreev bound state is clearly seen at
$\vare = \epsilon_\sm{B}$.
(Lower right): DOS at distance $x=\lambda/8$ from domain wall, the Andreev bound state
is slightly suppressed.
(Upper right): DOS at the maximum of order parameter, the Andreev bound state is
completely suppressed.
In addition to the broadened Andreev bound state at 
$\vare = \epsilon_\sm{B}$
there are singularities due to multiple Andreev reflections.
}
\end{figure}

Regions in which the order parameter changes sign, as in the LO phase, have 
an associated spectrum on low-energy quasiparticles, which in this case is also 
spin-polarized. The local quasiparticle density of states (LDOS) is easily calculated 
from the retarded Green's function $\whg^\sm{R}(\vR, \hat{\vp}; \vare)$, which is 
determined from Eq. (\ref{eq:eil}) using the self-consistently determined 
order parameter and analytic continuation to real energies from the upper half
plane: $i\vare_m \to \vare +i0^+$. The spatial and angle-resolved densities of states 
(DOS) for spin-up and spin-down excitations along the magnetic field are defined by
\be
N_{\uparrow(\downarrow)}(\vR, \hat{\vp}; \vare) = 
-{1\over \pi} \Im (g^\sm{R} 
\pm
\vg^\sm{R}\cdot\hat{\vb}) 
\,.
\ee  
In Figs.~\ref{fig:dos110sd}-\ref{fig:dos100ps} we show the evolution of
the LDOS and spatially averaged DOS for spin-up and spin-down excitations
for $T/T_c=0.15$
as a function of magnetic fields and position in space. The spin-down DOS 
are shown as solid lines, while the spin-up DOS are indicated by dashed lines.


Fig.~\ref{fig:dos110sd} shows the angle-averaged, spin-dependent LDOS
for a single domain along the nodal $\langle 110 \rangle$ direction. 
This configuration is in many ways equivalent to the scenario of the
order parameter suppression near a $\langle 110 \rangle$ interface in 
a clean $d$-wave superconductor, c.f. Barash et al.\cite{bar97}
There is a strong Andreev bound state at the position of the domain wall,  
arising from the change of the sign of the order parameter along 
quasiparticle trajectories crossing the domain wall (middle panel). 
In a magnetic field the Andreev bound state is no longer pinned to the Fermi
level, but is shifted to $\vare = \epsilon_\sm{B} = \mu B \; (-\mu B)$
for spin-up (spin-down) quasiparticles. As we move away from the domain wall the 
bound state decays and the bulk DOS is eventually restored (lower panel), 
albeit with weak resonances at $\vare = \pm \epsilon_\sm{B}$.

Andreev bound states and resonances are characteristic features of the FFLO state. 
They exist as a result of sign changes of the order parameter,
and are visible in the DOS for the periodic $\langle 110 \rangle$ stripe phases 
as well, see Fig.~\ref{fig:dos110ps}. The resonance broadens due to hybridization of the 
energy levels associated with adjacent domain walls. Consequently the sharp bound state for 
a single domain wall broadens into a band that fills the low energy region below
the maximum gap to the continuum spectrum. The quasiparticle resonance is visible as a broad 
peak centered at $\vare=\pm\epsilon_\sm{B}$, even in the spatially averaged DOS as shown in 
Fig.~\ref{fig:dos110av}. Note also the evolution of the spectrum with increasing magnetic field 
(from bottom to top panel). As the field increases the resonance broadens and the DOS approaches 
the normal-state value $N(\vare)/N_f = 1$, as is expected for a second-order transition to the
normal state.

Evidence of the FFLO state could be obtained from scanning tunneling microscopy by direct observation 
of the Andreev resonances organized along one-dimensional stripes, similar to the spectroscopic
observation of vortex arrays in type II superconductors.\cite{hes89f} Since the quasiparticle resonance 
spectrum and distribution in space depend on the magnetic field, these states should be easily 
differentiated 
from resonances arising from scattering by impurities or interfaces.\cite{sal96,pan99,pan03}
For the same reasons, point-contact spectroscopy, similar to recent measurements performed 
on CeCoIn$_5$,\cite{rou05,par05} but for magnetic fields in the FFLO phase, should be able 
to identify the characteristic signatures of the FFLO state.

In Fig.~\ref{fig:dos100ps} we show the DOS for stripes oriented along $\langle 100 \rangle$.
In contrast to the $\langle 110 \rangle$ orientation, there is a much richer spectrum
along the $\langle 100 \rangle$ direction. Additional singularities are visible at 
$\varepsilon/(2\pi T_c) \sim 0.02, 0.12, 0.28, 0.38$. These singularities originate from 
quasiparticles that propagate nearly parallel to the stripes. For these trajectories there is 
a broad dip in the order parameter amplitude that produces bound states due to multiple 
Andreev reflections from the "walls" of this ``order parameter well''.
These resonances generally occur at finite energy, in contrast to the topological bound states 
which are pinned to zero-energy in zero field. The bound states are further shifted by the field,
and the topological bound states - shifted by the Zeeman field - arising from trajectories crossing 
the stripes are also visible in the spectrum.

\begin{figure}[t]
\centerline{\includegraphics[width=8.0cm]{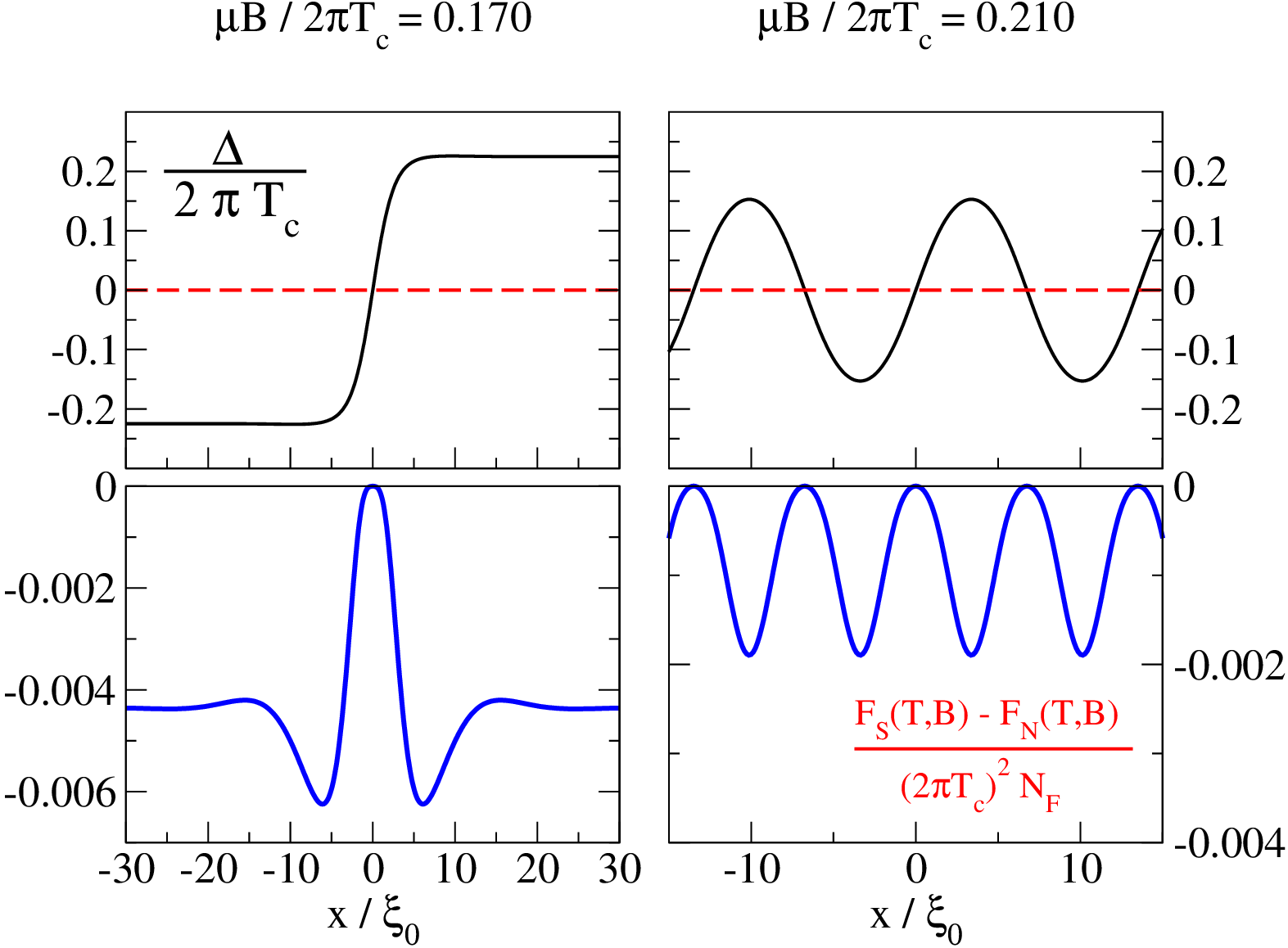}}
\caption{\label{fig:fe110} 
(Upper left): Profile of order parameter for a single domain wall along $\langle 110 \rangle$. 
(Lower left): Profile of free energy density $\Del f$ for single domain wall.
(Upper right): Profile of order parameter for periodic domain walls along $\langle 110 \rangle$. 
(Lower right): Profile of free energy density $\Del f$ for periodic domain walls.
}
\end{figure}

\begin{figure}[t]
\centerline{\includegraphics[width=8.0cm]{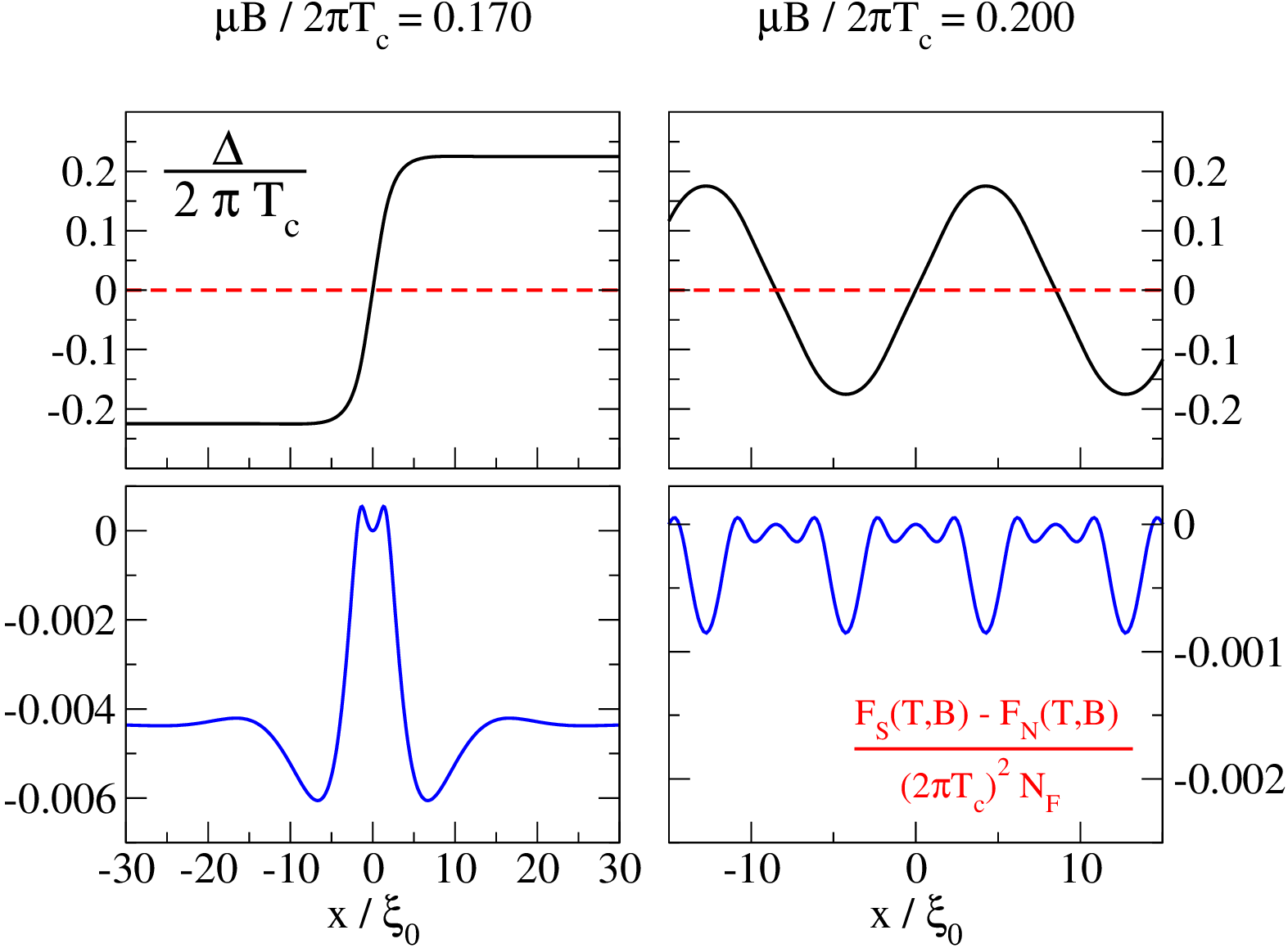}}
\caption{\label{fig:fe100} 
(Upper left): Profile of order parameter for a single domain wall along $\langle 100 \rangle$. 
(Lower left): Profile of free energy density $\Del f$ for single domain wall.
(Upper right): Profile of order parameter for periodic domain walls along $\langle 100 \rangle$. 
(Lower right): Profile of free energy density $\Del f$ for periodic domain walls.
}
\end{figure}

The spectrum for $\langle 100 \rangle$ stripes is similar to the LO state in $s$-wave 
superconductors because quasiparticles propagate along trajectories with similar order parameter 
profiles for directions perpendicular and parallel to the stripes. 
However, the anisotropy of the $d$-wave gap function in momentum space leads to more singularities 
below the maximum gap compared to the isotropic $s$-wave case. 

Finally, we note that Maki, Won and collaborators\cite{mak02,won04a,won04b} calculated the spin-averaged
DOS for the $\langle 100 \rangle$ striped FFLO state near the upper critical field at $T=0$. Their result 
for the spatially averaged DOS, expanded to second order in the order parameter, 
is qualitatively different from our self-consistently calculated DOS at $T/T_c = 0.15$
and $b=0.2$ (see Fig.~\ref{fig:dos100ps}); sufficiently so that a direct comparison is not possible.

\section{\label{sec:FE} Free energy density}

The bound state and resonance spectra associated with the FFLO phases 
shown in Figs.~\ref{fig:dos110ps} and \ref{fig:dos100ps} for the quasiparticle
density of states also lead to 
structure in the free energy densities. For the stripe orientation along
$\langle 110 \rangle$ see Fig.~\ref{fig:fe110},
and for $\langle 100 \rangle$ stripes see Fig.~\ref{fig:fe100}.
The free energy density exhibits additional structure for stripes oriented along 
$\langle 100 \rangle$ compared to $\langle 110 \rangle$.
At the domain walls the free energy density is that of the normal
state, while it is minimum where the magnitude of the order parameter 
reaches its maximum. The overshooting behavior of the free energy density
within a few coherence lengths away from the domain wall, as seen for a single 
domain wall in the lower left panels of Figs.~\ref{fig:fe110} and \ref{fig:fe100}, 
accounts for the energy gained compared to the uniform superconducting state.

The free energy density plots for a single domain wall shown in the lower
left panels of Figs.~\ref{fig:fe110} and \ref{fig:fe100} illustrate the subtle
differences in energy between order parameter modulations along 
$\langle 110 \rangle$ and $\langle 100 \rangle$ directions. Only a self-consistent
calculation of the nonuniform order parameter can resolve such detail.

\section{\label{sec:CLS} Conclusions}

We presented self-consistently calculated phase diagram of the 
FFLO state of a two-dimensional $d$-wave superconductor within
the quasiclassical theory. 
We found that the Larkin-Ovchinnikov state is favored 
over the Fulde-Ferrell state in two dimensions. The oscillations 
of the order parameter amplitude along the nodal $\langle 110 \rangle$ direction 
for the gap function $\cos 2\phi$ is stable over most of the FFLO phase diagram. 
However, for one-dimensional stripes the state with $\langle 100 \rangle$ orientation
is stable in the ultra-high $B$ region of the 
phase diagram and below $T^* \approx 0.06\, T_c$. Both $B_{c2}$ and 
$B_{c1}$ transitions are of second order. The order parameter vanishes continuously 
at the upper critical field $B_{c2}$. At the lower critical field $B_{c1}$
a single domain wall, similar to a Bloch domain wall, signals the onset
of the nonuniform FFLO state.

The local density of states calculations suggest that the Andreev resonance spectrum
may be used to identify experimentally the intrinsic structure of the FFLO phases.
In particular, the topological Andreev bound state is shifted from zero energy by an applied 
magnetic field. All the quasiparticle resonances live in regions where the order parameter 
is suppressed. Thus their spatial distribution depends on the wave length of the 
nonuniform modulations, which are controlled by the strength of the magnetic field. 
Besides the characteristic energy dependence of the resonances, their spatial 
periodicity in the FFLO state enables us to tell them apart from random impurities 
in scanning tunneling microscopy measurements.

Finally, we note that the phase diagram calculated for the model two-dimensional $d$-wave 
superconductor does not resemble the experimental phase diagram of CeCoIn$_5$. 
Thus, the identification of the observed phase transition in CeCoIn$_5$ with a FFLO
phase of a $d$-wave superconductor is either incorrect, or inconclusive and 
a more detailed model for the FFLO phases in this system needs to be implemented, e.g.,
a model with includes impurity scattering effects, Fermi liquid effects (e.g., exchange 
interactions), orbital magnetization effects as well as more realistic Fermi surface 
topology. All of these materials effects are known to result in 
changes to the order and the location of the phase transition lines. 

\section{Acknowledgments}

We like to thank C. Capan, T. L\"ofwander, M. Fogelstr\"om, I. Vekhter,
L. Bulaevskii, and A.V. Balatsky for many helpful discussions. 
ABV and JAS thank the CNLS and T-11, respectively, for their 
hospitality and CNLS's computing support.
This work was supported by the U.S. Department of Energy and funded through the
Los Alamos National Laboratory (JAS and MJG). ABV received
partial funding from the CNLS and the Louisiana Board of Regents.


\begin{thebibliography}{65}
\bibitem{footnote}
{We measure temperature and frequency in energy units and set $\hbar=k_B=1$.}
\end{thebibliography}
\end{document}